# A multivariate spatial interpolation of airborne γ-ray data using the geological constraints

E. Guastaldi[a], M. Baldoncini[c], G. P. Bezzon[d], C. Broggini[b], G. P. Buso[d], A. Caciolli[b], Carmignani L.[a], I. Callegari[a], T. Colonna[a], K. Dule[h], G. Fiorentini[c,d,e], M. Kaçeli Xhixha[g], F. Mantovani[c,e], G. Massa[a], R. Menegazzo[b], L. Mou[d], C. Rossi Alvarez[b], V. Strati[c], G. Xhixha[c,d,f], A. Zanon[d]

[a] CGT Center for GeoTechnologies, University of Siena, Via Vetri Vecchi, 34 - 52027 San Giovanni Valdarno, Arezzo, Italy.
[b] Istituto Nazionale di Fisica Nucleare (INFN), Padova Section, Via Marzolo 8 - 35131 Padova, Italy.
[c] Department of Physics and Earth Sciences, University of Ferrara, Via Saragat, 1 - 44100 Ferrara, Italy.
[d] Istituto Nazionale di Fisica Nucleare (INFN), Legnaro National Laboratory, Via dell'Università, 2 - 35020 Legnaro, Padova, Italy.
[e] Istituto Nazionale di Fisica Nucleare (INFN), Ferrara, Via Saragat, 1 - 44100 Ferrara, Italy.
[f] Faculty of Forestry Science, Agricultural University of Tirana, Kodër Kamëz - 1029 Tirana, Albania.
[g] University of Sassari, Botanical, Ecological and Geological Sciences Department, Piazza Università 21- 07100 Sassari, Italy.
[h] Faculty of Natural Sciences, University of Tirana, Bulevardi "Zogu I", Tirana, Albania.

## Abstract

In this paper we present maps of K, eU, and eTh abundances of Elba Island (Italy) obtained with a multivariate spatial interpolation of airborne γ-ray data using the constraints of the geologic map. The radiometric measurements were performed by a module of four NaI(Tl) crystals of 16 L mounted on an autogyro. We applied the collocated cokriging (CCoK) as a multivariate estimation method for interpolating the primary under-sampled airborne γ-ray data considering the well-sampled geological information as ancillary variables. A random number has been assigned to each of 73 geological formations identified in the geological map at scale 1:10,000. The non-dependency of the estimated results from the random numbering process has been tested for three distinct models. The experimental cross-semivariograms constructed for radioelement-geology couples show well-defined co-variability structures for both direct and crossed variograms. The high statistical correlations among K, eU, and eTh measurements are confirmed also by the same maximum distance of spatial autocorrelation. Combining the smoothing effects of probabilistic interpolator and the abrupt discontinuities of the geological map, the results show a distinct correlation between the geological formation and radioactivity content. The contour of Mt. Capanne pluton can be distinguished by high K, eU and eTh abundances, while different degrees of radioactivity content identify the tectonic units. A clear anomaly of high K content in the Mt. Calamita promontory confirms the presence of felsic dykes and hydrothermal veins not reported in our geological map. Although we assign a unique number to each geological formation, the method shows that the internal variability of the radiometric data is not biased by the multivariate interpolation.

50

51    **1. Introduction**

52

53    Airborne γ-ray spectrometry (AGRS) is a fruitful method for mapping natural
54    radioactivity, both in geoscience studies and for purposes of emergency response. One of
55    the principal advantages of AGRS is that it is highly appropriate for large scale
56    geological and environmental surveys (**Minty, 2011**; **Sanderson et al., 2004**; **Rybach et**
57    **al., 2001**; **Bierwirth & Brodie, 2008**). Typically, the AGRS system is composed of four
58    4 L NaI(Tl) detectors mounted on an aircraft. For fixed conditions of flight a challenge is
59    to increase the amount of geological information, developing dedicated algorithms for
60    data analysis and spatial interpolation. The full spectrum analysis (FSA) with the non-
61    negative least squares (NNLS) constraint (**Caciolli et al., 2012**) and noise-adjusted
62    singular value decomposition (NASVD) analysis (**Minty & McFadden, 1998**) introduces
63    notable results oriented to improve the quality of the radiometric data. On the other hand,
64    the multivariate interpolation has the great potential to combine γ-ray data with the
65    preexisting information contained in geological maps for capturing the geological local
66    variability.

67

68    Elba Island (Italy) is a suitable site for testing a multivariate interpolation applied to
69    AGRS data because of its high lithological variability, excellent exposure of outcropping
70    rocks and detailed geological map. In multivariate statistical analysis, different pieces of information about the particular characteristics of a variable of interest may be better predicted by combining them with other interrelated ancillary information into a single optimized prediction model. This approach improves the results of the spatial interpolation of environmental variables. However, sometimes primary and ancillary variables are sampled by different supports, measured on different scales, and organized in different sampling schemes, which makes spatial prediction more difficult.

In this study the collocated cokriging (CCoK) was used in a non-conventional way for dealing with the primary (AGRS data) and secondary (geological data) variables when the variable of interest has been sampled at a few locations and the secondary variable has been extensively sampled. Using this approach, we provide the map of natural radioactivity of Elba Island.

## 2. Instruments and methods

*2.1. Geological setting*

Elba is the biggest island of the Tuscan Archipelago and is located in the northern part of the Tyrrhenian Sea, between Italy and Corsica Island (France). It is one of the westernmost outcrop of the Northern Apennines mountain chain (**Figure 1**).

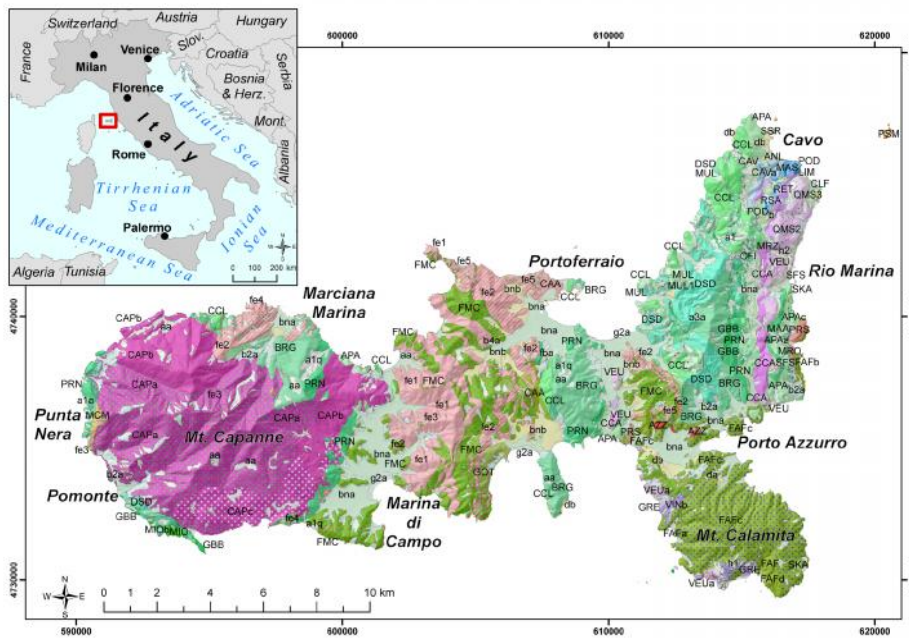

**Figure 1**. Geological map of Elba Island (taken from the Geological Map of Tuscany region realized at scale 1:10,000, see **CGT, 2011**): the western sector is mainly characterized by intrusive igneous rocks (magenta), the central and eastern sectors are characterized by a wide lithological variation (green, purple, and pink), while the southeastern outcrop is constituted almost exclusively of metamorphic rocks (Mt. Calamita). For the legend of the geologic map, see http://www.geologiatoscana.unisi.it. The coordinate system is UTM WGS84 Zone 32 North.

The geological distinctive features of this island are linked to its complex stack of tectonic units and the well-known Fe-rich ores, as well as the well-exposed interactions between Neogene magmatic intrusions and tectonics (**Trevisan, 1950**; **Bortolotti et al., 2001**; **Dini et al., 2002**; **Musumeci & Vaselli, 2012**). The structure of Elba Island consists of thrust sheets stacked during the late Oligocene to middle Miocene northern Apennines deformation. Thrust sheets are cross-cut by late Miocene extensional faults (**Keller & Coward 1996**; **Bortolotti et al., 2001**; **Smith et al., 2011**).

The tectonics of Elba Island is composed of a structural pile of five main units called by **Trevisan (1950)** as "Complexes" and hereafter called "Complexes of Trevisan" (TC): the lowermost three belong to the Tuscan Domain, whereas the uppermost two are related to the Ligurian Domain. **Bortolotti et al. (2001)** performed 1:10,000 mapping of central-eastern Elba and proposed a new stratigraphic and tectonic model in which the five TC were reinterpreted and renamed. TCs are shortly described below.

The Porto Azzurro Unit (TC I) (Mt. Calamita Unit Auct.) consists of Paleozoic micaschists, phyllites, and quartzites with local amphibolitic horizons, as well as Triassic-Hettangian metasiliciclastics and metacarbonates. Recently **Musumeci et al. (2011)** point out Early Carboniferous age for the Calamita Schist by means of U-Pb and $^{40}$Ar-$^{39}$Ar radioisotopic data. In particular, in the Porto Azzurro area and the eastern side of Mt. Calamita, the micaschists are typically crosscut by the aplitic and microgranitic dykes that swarm from La Serra-Porto Azzurro monzogranitic pluton (5.1-6.2 Ma, **Dini et al., 2010** and references therein). Magnetic activities have produced thermometamorphic imprints in the host rocks (**Garfagnoli et al., 2005**; **Musumeci & Vaselli, 2012**).

The Ortano Unit (lower part of TC II) includes metavolcanics, metasandstone, white quartzites and minor phyllites. The Acquadolce Unit (upper part of TC II) is composed of locally dolomitic massive marbles, grading upwards to calcschists (**Pandeli et al., 2001**). This lithology is capped by a thick siliciclastic succession. Ortano and Acquadolce units experienced late Miocene contact metamorphism under low to medium metamorphic grade conditions (**Duranti et al., 1992**; **Musumeci & Vaselli 2012**).

The Monticiano-Roccastrada Unit (lower part of TC III) includes basal fossiliferous graphitic metasediments of the Late Carboniferous-Early Permian, unconformably overlain by the detrital Verrucano succession (Middle-Late Triassic) (**Bortolotti et al., 2001**). The Tuscan Nappe Unit (central part of TC III) is represented by calcareous-dolomitic breccias and overlying carbonatic outcrops northwards. Most of Grassera Unit (upper part of TC III) is composed of varicolored slates and siltstones with rare metalimestone or meta-chert intercalations; basal calcschists also occur.

The Ophiolitic Unit (TC IV) is composed of several minor thrust sheets or tectonic sub-units, which are characterized by serpentinites, ophicalcites, Mg-gabbros, and Jurassic-Lower Cretaceous sedimentary cover (**Bortolotti et al., 2001**).

The Paleogene Flysch Unit (lower part of TC V) mainly consists of shales, marls with limestone, sandstone, and ophiolitic breccia intercalations including fossils of the Paleocene-Eocene age. The Lower-Upper Cretaceous Flysch Unit (upper part of TC V) consists of basal shales and varicolored shales. These lithologies vertically pass to turbiditic siliciclastic sandstones and conglomerates, which in turn alternate with marlstones and marly limestones. Both Flysch Units were intruded by aplitic and porphyritic dykes and laccoliths approximately 7-8 Ma ago (**Dini et al., 2002**).

The geological structure of the island allows a nearly complete representation of lithologies present in the Northern Apennines mountain chain (**Figure 1**). This feature makes Elba Island a complex system in terms of both geological formations and lithologies. Therefore, it is a formidable research site for applying a multivariate interpolation of radiometric data in relationship to lithological properties.

*2.2. Experimental setup, survey, and data*

The AGRS system is a modular instrument composed of four NaI(Tl) detectors (10 × 10 × 40 cm each) with a total volume of 16 L mounted on an autogyro (**Figure 2**). The system is further equipped with a 1 L "upward-looking" NaI(Tl) detector, partially shielded from the ground radiation and used to account for atmospheric radon. Other auxiliary instruments, including the GPS antenna and pressure and temperature sensors, are used to record the position of the AGRS system and to measure the height above the ground using the Laplace formula (**IAEA, 1991**).

168
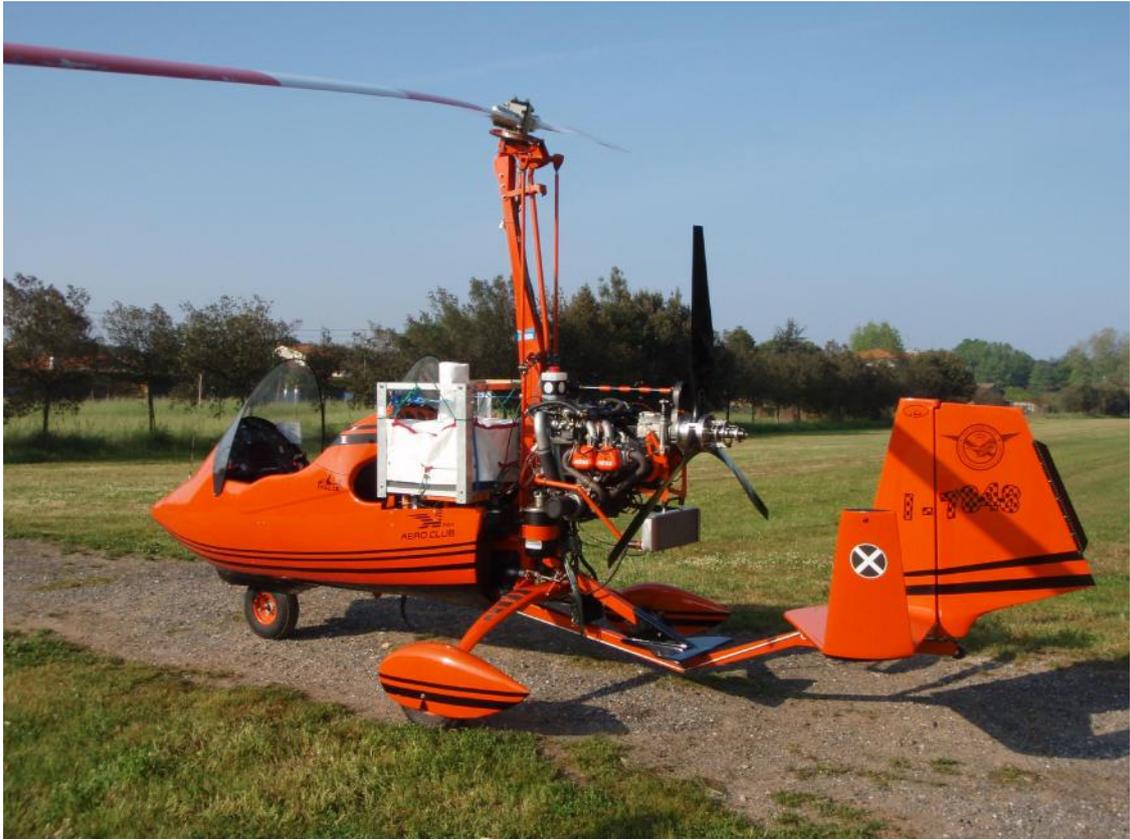

169
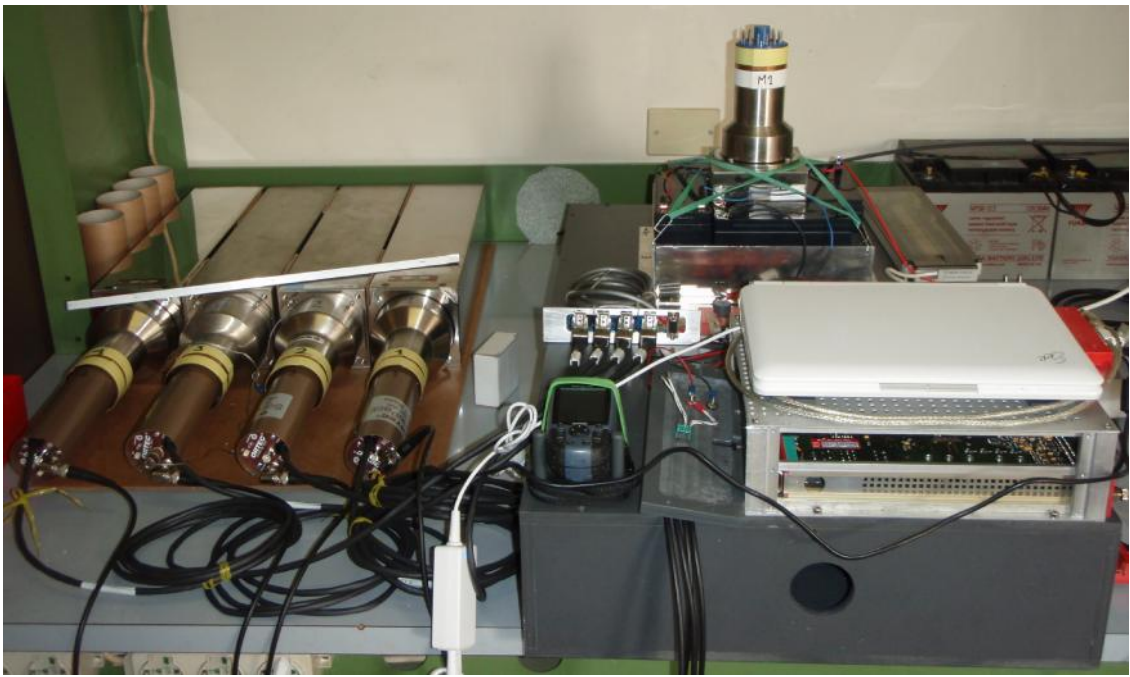

170 **Figure 2**. The airborne γ-ray setup (a) mounted on the autogyro (b). The main detector system is inserted in
171 the box under the "upward-looking" detector, which is placed behind the laptop.

As a survey strategy, we planned to be as perpendicular as possible to the main N-S strike of the geological structures of the area (**Figure 1**). The flight lines were designed in a spiral structure, constrained by the morphology of the terrain (elevations 0÷1010 m a.m.s.l.), starting from the shore and following the heights of the mountains in the counterclock direction (**Figures 6**, **7**, **8**). The unique region not properly covered by the airborne γ-ray survey is the top of Capanne Mt., because of the cloudy weather conditions. Averaging the flight altitudes recorded every two seconds we have 140 ± 50 m (standard deviation). The survey parameters were designed for a cruise speed of approximately 100 km/h, with space lines at most 500 m from one another. For our flight conditions, the detection system is able to measure the signal (97%) coming from a spot area of approximately 600 m radius, even if 90% comes from the half of this radius. In this study, the effect of attenuation of the signal from the biomass (**Schetselaar et al., 2000**; **Carroll & Carroll, 1989**) was neglected since Elba Island is covered by a large extension of rock outcrops and scattered vegetation of Mediterranean scrub.

The signal is acquired in list mode (event by event) using an integrated electronic module with four independent signal-processing channels and then analyzed offline in 10s intervals. This time interval is chosen such as to optimize the loss in spatial resolution and to reduce the statistical uncertainty to less than 10%. The γ-spectra are calibrated and analyzed using the full spectrum analysis with non-negative least squares (FSA-NNLS) approach as described in **Caciolli et al. (2012)**. According to the FSA method the spectrum acquired during the offline analysis is fitted as a linear combination of the fundamental spectra derived for each radioelement and for background from the calibration process. The abundances are determined applying the non-negative least squares to minimize the $\chi^2$: the NNLS algorithm reduces the presence of non-physical results, which can lead to systematic errors (Caciolli et al., 2012).

Several corrections are applied to the signal measured at different flight altitudes to determine the concentrations of K, eU (equivalent uranium) and eTh (equivalent thorium) at the ground: a) aircraft and cosmic background correction; b) topology correction; c) flying altitude and height correction and d) atmospheric radon correction. The dead time correction was found to be negligible due to relatively low count rates measured during the flight. The background correction is taken into account during the calibration process where the fundamental spectra of the background due to the aircraft and cosmic radiation is estimated. The numeric regional topographic map at 1:10,000 scale of the ground surface has been accounted for the digital elevation model, which has a 10 m spatial resolution. The effects of the steep Elba Island's topography (ranging between 0 m to 1010 m a.m.s.l.) are corrected following the method described in Schwarz et al. (1992). Finally, to compute the concentration at the ground surface, the signal is further corrected by an empirical factor obtained by measuring the signal at several altitudes over a flat surface well characterized by ground measurements. The altitude and topography corrections introduce a total systematic uncertainty on the order of 10% in the final results.

Further corrections are required for eU concentration because the signal coming from ground uranium is increased by the radon gas in the air. It is evaluated by using the method of the "upward-looking" detector, following the procedure described in **IAEA (1991)**. The atmospheric radon concentration is estimated by analyzing the spectrum acquired with the "upward-looking" detector, which is calibrated by flying over the Tyrrhenian sea at the beginning and the end of the survey. The radon concentration has been calculated for each time interval and was almost stable during the entire flight (0.2 ± 0.1 μg/g). Since the ground abundance of eU varies from 0.2 μg/g up to 28.0 μg/g over all of Elba Island, the uncertainty concerning the atmospheric radon subtraction for each single measurement varies from 2% up to 100%: indicatively in average the relative uncertainty was 23%.

The relative uncertainties for K, eU, and eTh abundances[1] in the final results are summarized in **Table 1**. The systematic relative uncertainties are estimated by combining the contributions from the altitude and topography corrections and the calibration process. We emphasize that the data used as input in the CCoK interpolation are taken into account without experimental uncertainties and that their positions are related to the center of the spot area.

**Table 1**. Experimental relative uncertainties for the measured abundances of K, eU, and eTh.

| Radionuclide | Statistical | Systematic |
|---|---|---|
| K | 7% | 14% |
| eU | 8% | ~ 30%[a] |
| eTh | 8% | 15% |

---

[1] The activity concentrations of 1 μg/g U (Th) corresponds to 12.35 (4.06) Bq/kg and 1% K corresponds to 313 Bq/kg.

[a] includes the uncertainty related to atmospheric radon correction.

*2.3. Geostatistical data analysis*

Geostatistics involves spatial datasets, predicting distributions that characterize the coregionalization between the variables. The CCoK is a special case of cokriging wherein a secondary variable is available at all prediction locations is used to estimate a primary under-sampled variable, restricting the secondary variable search to a local neighborhood. Frequently, the primary and ancillary (secondary) variables are sampled by different supports, measured on different scales, and organized in different sampling schemes, making the spatial prediction more complex. The integration of data that may differ in terms of type, reliability, and scale has been studied in several works. In **Babak & Deutsch (2009)**, for instance, this approach is adopted using dense 3D seismic data and test data for an improved characterization of reservoir heterogeneity.

This approach is also used for mapping soil organic matter (**Pei et al., 2010**), rainfall, or temperature over a territory (**Goovaerts, 1999**; **Hudson & Wackernagel, 1994**); ground based radiometry data (**Atkinson et al., 1992**); estimating environmental variables, such as pollutants or water tables (**Guastaldi & Del Frate, 2012**; **Desbarats et al., 2002**; **Hoeksema et al., 1989**); and mapping geogenic radon gas in soil (**Buttafuoco et al., 2010**). To date, this method has not been applied to airborne γ-ray measurements integrated with geological data. A multivariate technique for interpolating airborne γ-ray data on the basis of the geological map information is desirable.

261 We used the collocated cokriging as a multivariate estimation method for the
262 interpolation of primary under-sampled airborne γ-ray data using a constraint based on
263 the secondary well-sampled geological information. This section briefly describes the
264 theoretical background of CCoK interpolation and its application to airborne γ-ray data
265 using geological constraints.

266

267 *2.3.1. Collocated cokriging: theoretical background*

268

269 Geostatistical interpolation algorithms construct probability distributions that characterize
270 the present uncertainty by the coregionalization among variables (**Wackernagel, 2003**).
271 The CCoK is an interpolation method widely used when applying a linear
272 coregionalization model (LCM) to a primary under-sampled variable $Z_1(x)$ and a
273 secondary widely sampled variable $Z_2(x)$ continuously known at all grid nodes
274 (**Goovaerts, 1997**).

275

276 **Xu et al. (1992)** advanced a definition in which the neighborhood of the auxiliary
277 variable $Z_2(x)$ is arbitrarily reduced to the target estimation location $x_0$ only. They
278 formulated CCoK as a simple cokriging linked to the covariance structure (**Chiles &
279 Delfiner, 1999**):

280

281 $\rho_{12}(h) = \rho_{12}(0)\rho_{11}(h)$ \hfill (1)

282

283 where $\rho_{11}(h)$ is the correlogram of the primary variable $Z_1(h)$ and $\rho_{12}(h)$ is the cross-
284 correlogram, which quantifies the spatial correlation between the primary ($Z_1$) and the
285 secondary ($Z_2$) data at a distance $h$.

286 Assuming $Z_1(x)$ to be known, the value of the primary variable $Z_1$ at target location $x_0$
287 is independent of the value of the secondary variable $Z_2$ if $Z_1$ and $Z_2$ have a mean of
288 zero and a variance of one. In this case, which is called a "Markov-type" model, the cross
289 covariance functions are proportional to the covariance structure of the primary variable
290 (**Xu et al., 1992**; **Almeida & Journel, 1994**). A strictly CCoK estimator $Z_{1_{CCoK}}^{**}$ at target
291 location $x_0$ depends on both the linear regression of the primary variable $Z_1$ and the
292 simple kriging variance $\sigma_{SK}^2$, for $\rho = \rho_{12}(0)$ as follows (**Chiles & Delfiner, 1999**):

$$Z_{1_{CCoK}}^{**}(x_0) = \frac{(1-\rho^2)Z_1^*(x_0) + \sigma_{SK}^2 \rho Z_2(x_0)}{(1-\rho^2) + \rho^2 \sigma_{SK}^2} \quad (2)$$

296 where $Z_1^*$ is the kriging estimation of $Z_1$ at the target location $x_0$ and the accuracy of the
297 CCoK estimation is given by

$$\sigma_{CCoK}^2 = \sigma_{SK}^2 \frac{(1-\rho^2)}{(1-\rho^2) + \rho^2 \sigma_{SK}^2} \quad (3)$$

301 *2.3.2. Interpolating airborne γ-ray data on geological constraints*

In our study, we used the CCoK as a multivariate estimation method for the interpolation of airborne γ-ray data using the geological map information. The primary variable $Z_1(x)$ refers to the discrete distribution of the natural abundances of K, eU, or eTh (equivalent thorium) measured via airborne γ-ray spectrometry, whereas the secondary variable $Z_2(x)$ refers to the continuous distribution of the geological formations (i.e., the geological map). In this work, these two sets of information are independent of one another. The data gained through airborne γ-ray spectrometry define a radiometric spatial dataset integrating the sample point positions with the natural abundances of K (%), eU (μg/g), and eTh (μg/g), together with their respective uncertainties.

The geological map at a 1:10,000 scale (**CGT, 2011**), obtained from a geological field survey, covers the entire area in detail. Moreover, the geological map lists 73 different geological formations, defining in this way a categorical variable. For such a large number of variables, the approach based on categorical variables (**Hengl et al., 2007**; **Pardo-Iguzquiza & Dowd, 2005**; **Goovaerts, 1997**; **Rossi et al., 1994**; **Bierkens & Burrough, 1993**; **Journel, 1986**) requires a long time for processing and interpretation. Therefore, we had to consider the geological qualitative (categorical) map as a quasi-quantitative constraining variable. In order to study the frequency of sampling we sorted in alphabetical ascending order the geological formations names and assigned to each one a progressive number. We rearranged the frequencies for obtaining normal distributions of the secondary variable (geology). As we show in the following section, this procedure does not affect the final interpolation results. Thus, we spatially conjoined the airborne γ-ray measures to the geological map. This migration of geological data from the continuous grid (the geological map) to the sample points (the airborne γ-ray measuring locations) is performed to yield a multivariate point dataset to be interpolated by CCoK. As shown in **Table 2**, K (%) and eTh (μg/g) abundances have a quasi-Gaussian distributions, whereas eU (μg/g) abundance distribution tends to be positively skewed. The linear correlation is high between pairs of abundance variables (**Figure 3**). Based on the previous assumptions, the linear correlation coefficient between radioactivity measures and values arbitrary assigned to geological formations is meaningless.

**Table 2**. Descriptive statistical parameters of airborne γ-ray data.

| Parameter | K (%) | eU (μg/g) | eTh (μg/g) |
|---|---|---|---|
| Count | 806 | 805 | 807 |
| Minimum | 0.2 | 0.2 | 0.03 |
| Maximum | 4.8 | 28.0 | 34.0 |
| Mean | 1.9 | 6.4 | 11.1 |
| Std. Dev. | 0.9 | 4.4 | 5.9 |
| Variance | 0.8 | 19.7 | 35.2 |
| Variation Coeff. | 0.5 | 0.7 | 0.5 |
| Skewness | 0.2 | 1.3 | 0.5 |

The CCoK interpolation models, both for the direct spatial correlation and the cross-correlation of these regionalized variables, were obtained by calculating experimental semi-variograms (ESV) and experimental cross-semivariograms (X-ESV), and interpreting the models by taking into account factors conditioning the spatial distribution of these regionalized variables. The distributions of radioelements of our dataset show a positive skewness of 0.2, 0.5 and 1.3 for K, eTh and eU respectively (**Table 2**). In the case of skewness values less than 1, several authors (**Webster & Oliver 2001**; **Rivoirard 2001**) suggest to not perform any normal transformation of the data. Considering that the measurement of eU is contaminated by radon, which increases the experimental uncertainty, we considered redundant any refinement of data processing. In addition, supported by well-structured ESVs and X-ESVs for the raw datasets, we didn't perform any normal transformation for K, eU and eTh.

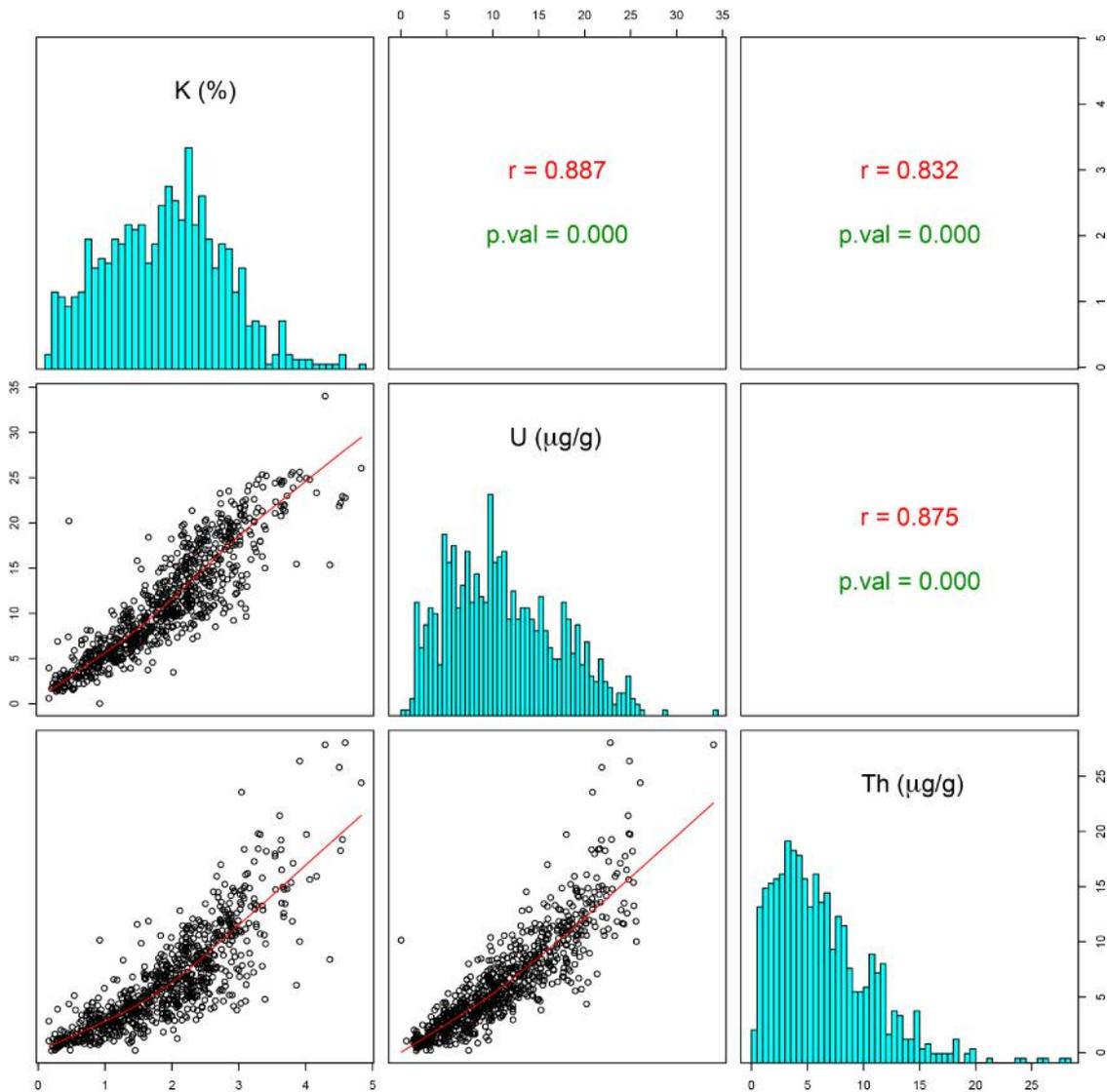

**Figure 3**. Correlation matrix of abundance variables: the lower panel shows the bi-variate scatter plots for each pair of variables and the robust locally weighted regression (**Cleveland, 1979**), red line; cells on the matrix diagonal show the univariate distributions of abundances; the upper panel shows both Pearson's

linear correlation coefficient value for each bivariate distribution and the statistical significance testing scores (p-value) for each correlation test.

The directional X-ESVs show erratic behavior. Therefore, we modeled the experimental co-variability as isotropic, and an omnidirectional LCM has been fitted using a trial-and-error procedure. As shown in **Table 3**, the Gaussian distribution has the mean of standardized errors equal to zero and the variance of standardized errors equal to unity, which allows us to use a cross-validation method. We double-checked the quality of the model (**Clark & Harper, 2000**; **Goovaerts, 1997**; **Isaaks & Srivastava, 1989**) by comparing the errors made in estimating airborne γ-ray measures at sample locations with the theoretical standard Gaussian distribution.

Each group of variables shows the same spatial variability of the geology in the coregionalization matrices because the same parametric variable is still used for all models in the estimation of abundance distribution maps of radioactive elements (**Table 3**). The result shows a well-structured spherical variability for all groups of variables (**Figure 4**).

**Table 3**. Parameters of linear coregionalization models fitted on omnidirectional variograms calculated with 8 lags of 200 m: groups of primary (radionuclides) and secondary variables; number and types of systems of functions fitted on experimental variograms; range distances for each system of function; matrices of each structure of variability of linear coregionalization model (LCM) fitted for different groups (model values for EVSs in each matrix diagonal cells, model values for XESVs in lower left panel of each matrix; variability values of the parametric geology, in the right column, are unitless); cross-validation

377 results of the fitted LCM (only the primary variables scores are listed; MSE: mean of standardized errors;
378 VSE: variance of standardized errors) for all groups of variables.

| Group of variables | Number and Type of Structures of variability | | Range (m) | LCM matrices | | Cross-validation | |
|---|---|---|---|---|---|---|---|
| | | | | | | MSE | VES |
| K & geology | 1 | Nugget Effect Model | - | 0.01 %$^2$ | - | -0.0016 | 0.68 |
| | | | | 0.3 %$^2$ | 15 | | |
| | 2 | Spherical Model | 400 | 0.1 %$^2$ | - | | |
| | | | | -0.6 %$^2$ | 87 | | |
| | 3 | Spherical Model | 1500 | 0.3 %$^2$ | - | | |
| | | | | -1.2 %$^2$ | 105 | | |
| eU & geology | 1 | Nugget Effect Model | - | 2.5 µg/g$^2$ | - | -0.00016 | 0.73 |
| | | | | 0.1 µg/g$^2$ | 87 | | |
| | 2 | Spherical Model | 1500 | 5.7 µg/g$^2$ | - | | |
| | | | | -5.7 µg/g$^2$ | 120 | | |
| eTh & geology | 1 | Nugget Effect Model | - | 0.4 µg/g$^2$ | - | -0.0008 | 0.65 |
| | | | | -0.1 µg/g$^2$ | 15 | | |
| | 2 | Spherical Model | 400 | 2.1 µg/g$^2$ | - | | |
| | | | | -0.4 µg/g$^2$ | 87 | | |
| | 3 | Spherical Model | 1500 | 11.2 µg/g$^2$ | - | | |
| | | | | -10.6 µg/g$^2$ | 105 | | |

379

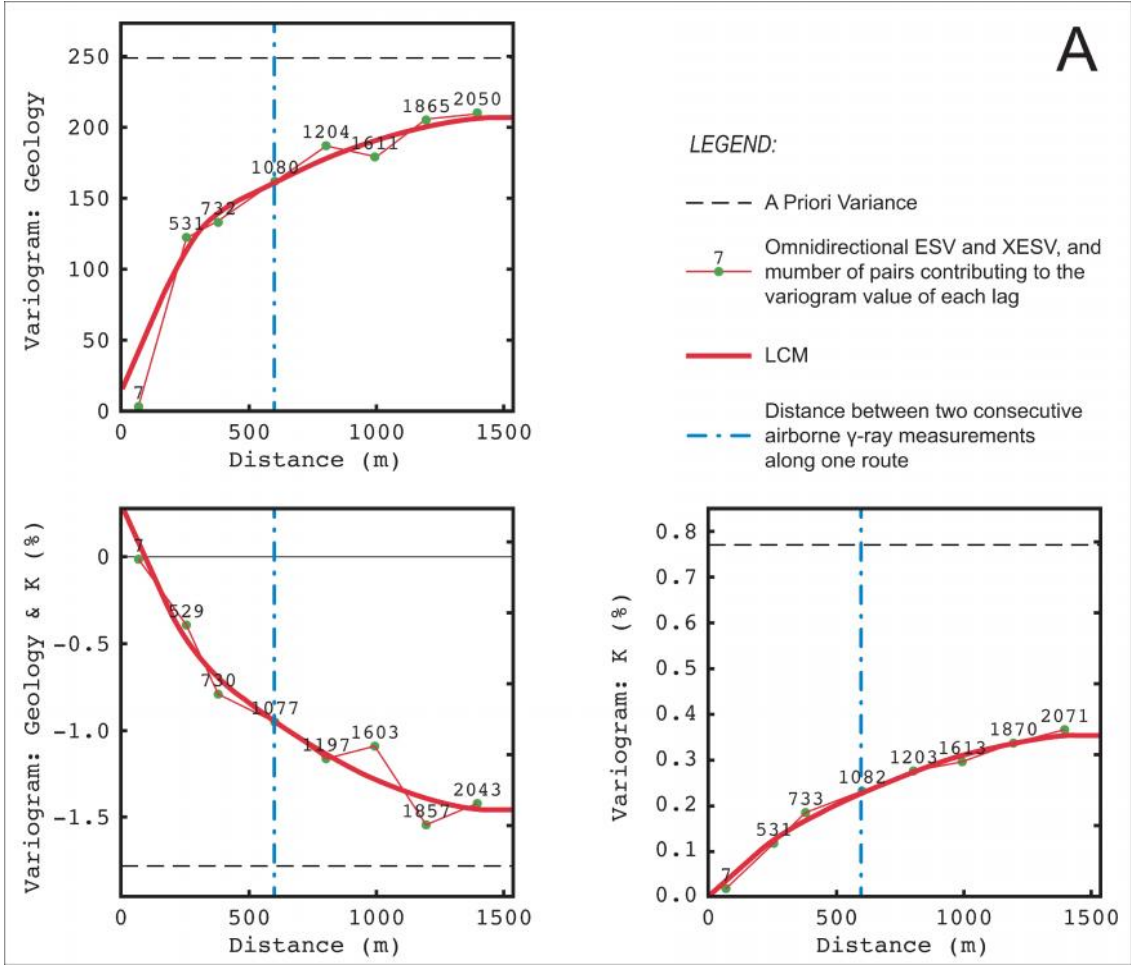

380

381 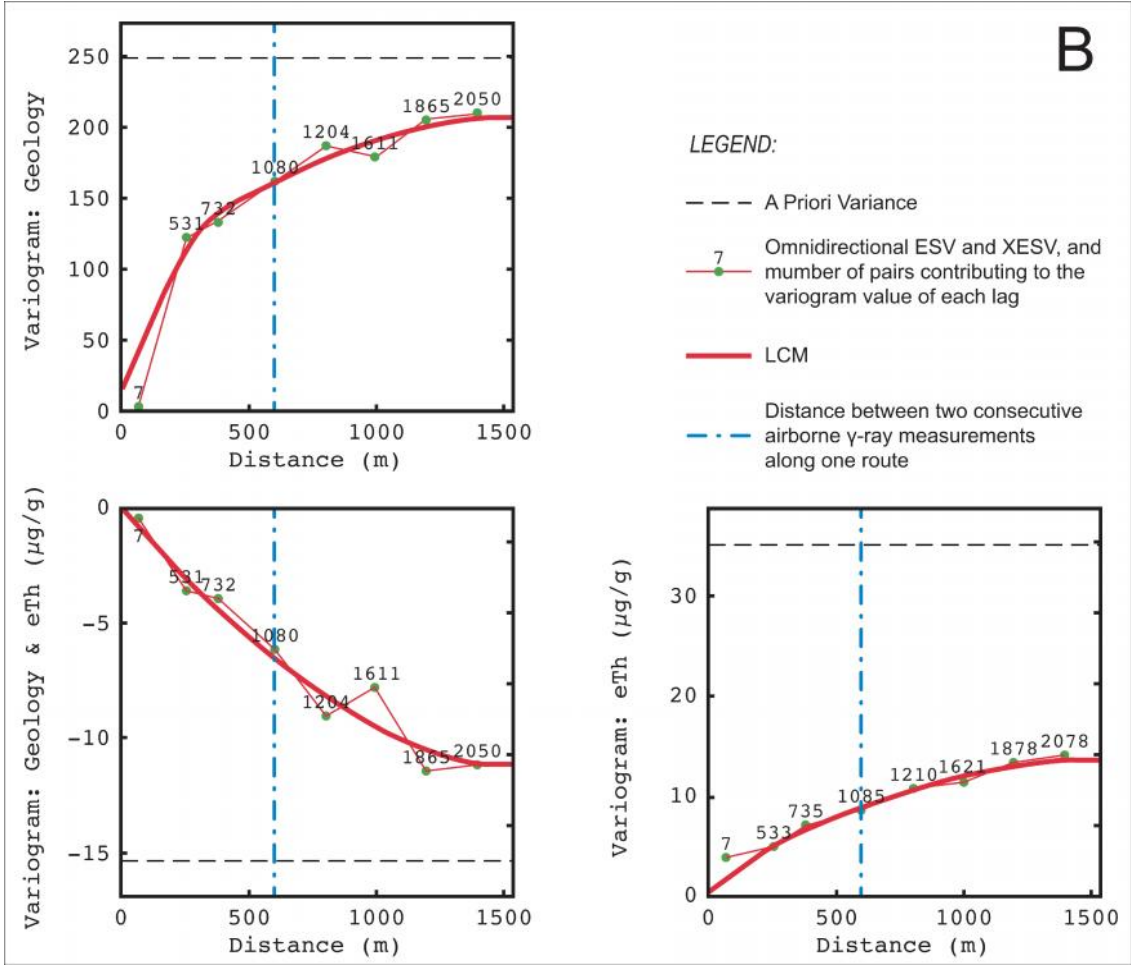

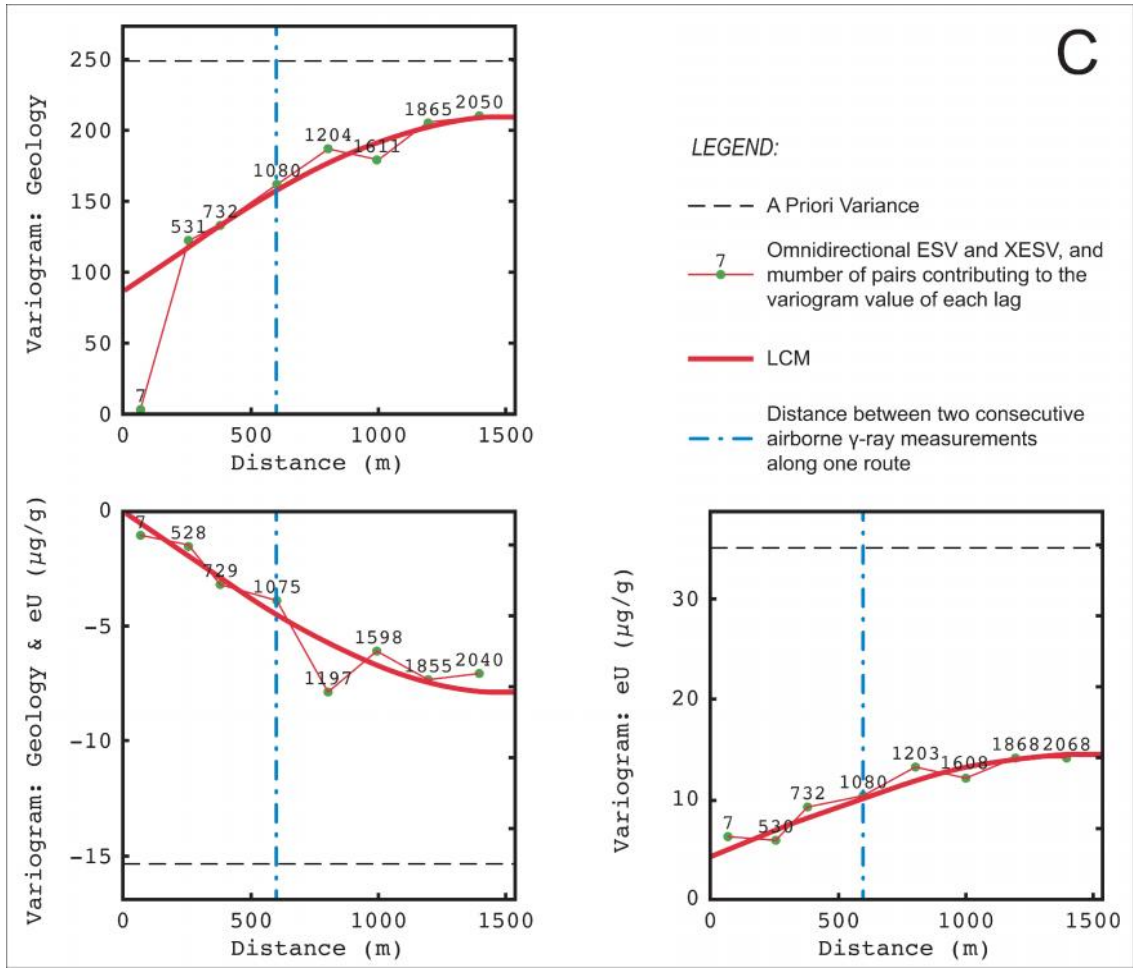

**Figure 4**. Omnidirectional Linear Coregionalization Model fitted for the experimental semi-variograms (ESV, on diagonal cells of the matrix) and cross-semivariograms (XESV lower left corner cell) for all groups of radionuclides and parametric geology: (a) Geology and K; (b) Geology and eTh; (c) Geology and eU.

## 3. Results and discussion

On the 3$^{rd}$ of June, 2010, the autogyro flew over Elba Island (224 km$^2$): during approximately two hours of flight, the ARGS system collected 807 radiometric data with

392    an average spot area of approximately 0.25 km$^2$ (source of 90% of the signal). The
393    average altitude of the flight was 140 ± 50 m.
394
395    Performing the post-processing described in **section 2.2**, we associated homogenous K,
396    eU, and eTh abundances to each spot area. Considering that 96% of the total 2574
397    geological polygons covering the surface of Elba Island have an area less than 0.25 km$^2$,
398    we observe that many of the airborne γ-ray measurements refer to the contributions
399    coming from several geological formations with different lithological compositions.
400    However, these polygons cover only 25% of the surface of Elba Island. The high density
401    of radioactivity data and the highly refined geological map allowed to construct a well
402    tested LCM: the cross-validation results are shown **Table 3**. Based on this consistent
403    framework, the multivariate analysis produced data characterized by a good assessment
404    of spatial co-variability. According to the flight plan, the autogyro crossed its own route
405    resulting in a very low variability in the first lags of the omnidirectional co-
406    regionalization model (e.g. ESV of K in **Figure 4 (a)**). The ESV models referred to
407    AGRS measurements show regular structures with low variability at small distances and
408    generally higher variability at the spherical parts. Indeed, the nugget effect of K
409    abundance contributes almost 2% of the total amount of spatial variability, providing
410    evidence of autocorrelation. The same features are found for the eTh and eU abundances,
411    whose variances at small distances contribute 3% and 30% of the total spatial variation,
412    respectively.
413

Moreover, we notice a low spatial variability below 600 m (indicating the spot area radius, indicated by the blue dashed line in **Figure 4**), which corresponds to data obtained by partially overlapping spot areas. The maximum distance of spatial autocorrelation for K, eU, and eTh is 1500 m (**Table 3**), this also due to their high statistical correlation (**Figure 3**). These features reconstructed the spatial resolution of the AGRS survey, confirming the consistency of the model and the AGRS data.

The variability of the parametric geology variogram at small distances show a weak variability discontinuity at lag $h = 0$, i.e., a nugget effect. This contributes almost 50% of the total spatial variability together with the first range of autocorrelation found at 400 m. This due to either the random values assigned to the categories of the geological map, where a significant difference can be found between the sample values of two adjacent geological formations or in the 600 m spot area radius (**Figure 4**).

The X-ESVs constructed for radioelement-geology couples generally show well-defined co-variability structures. Indeed, both the spherical components of the model are well structured and the contribution of the random part of the variability is always minimized (**Figure 4**). Therefore, we conclude that these choices ensure the consistency of the results achieved by the CCoK multivariate interpolator.

The estimated maps of the K, eTh, and eU abundances are shown in **Figures 6**, **7**, and **8**. These maps are calculated with a high spatial resolution (pixel size 10 m x 10 m) in accordance with the choice of the geological map at scale 1:10,000. We also report the accuracy of the estimations in terms of the variance, normalized respect to the estimated values of the abundances (normalized standard deviation, NSD). The percentage uncertainties of the abundances are higher when the absolute measures are smaller, with average NSDs of 27%, 28%, and 29% for K, eU, and eTh, respectively.

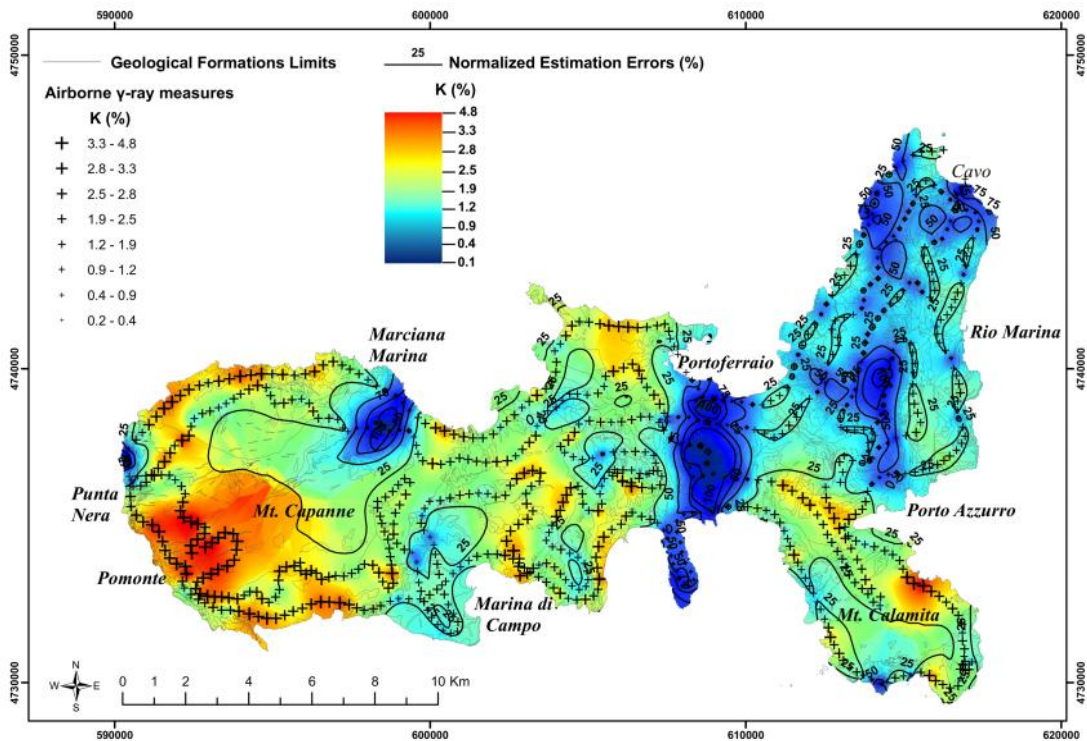

**Figure 5**. Estimation map of K (%) abundance and normalized estimation errors.

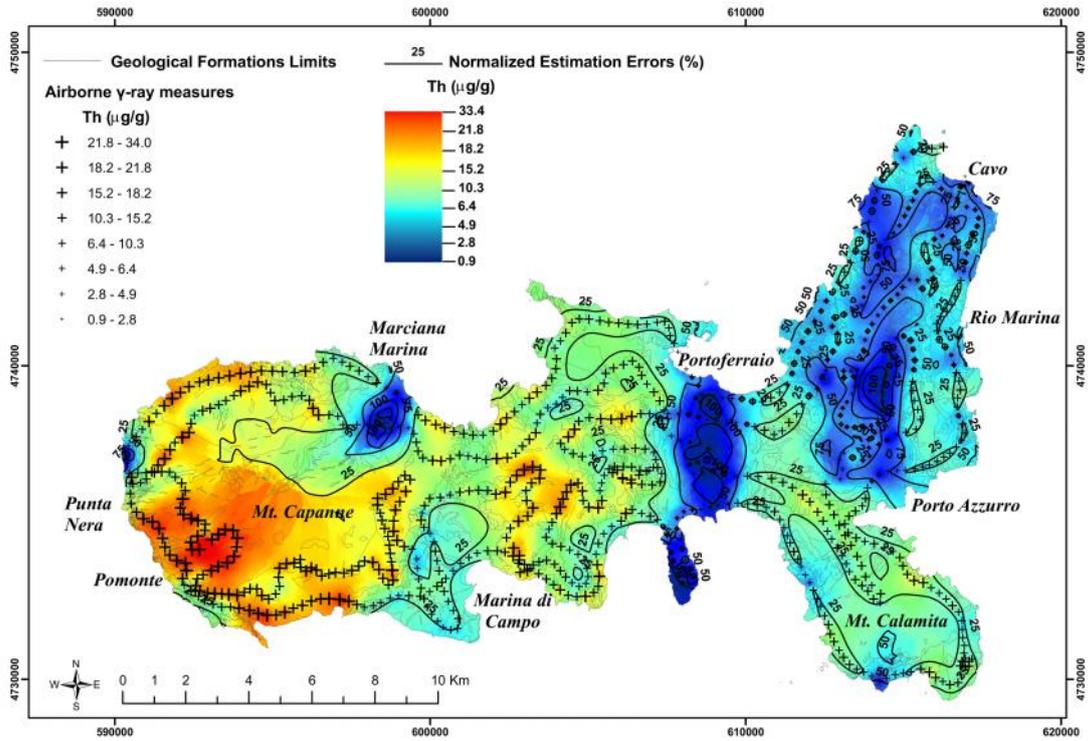

Figure 6. Estimation map of eTh (μg/g) abundance and normalized estimation errors.

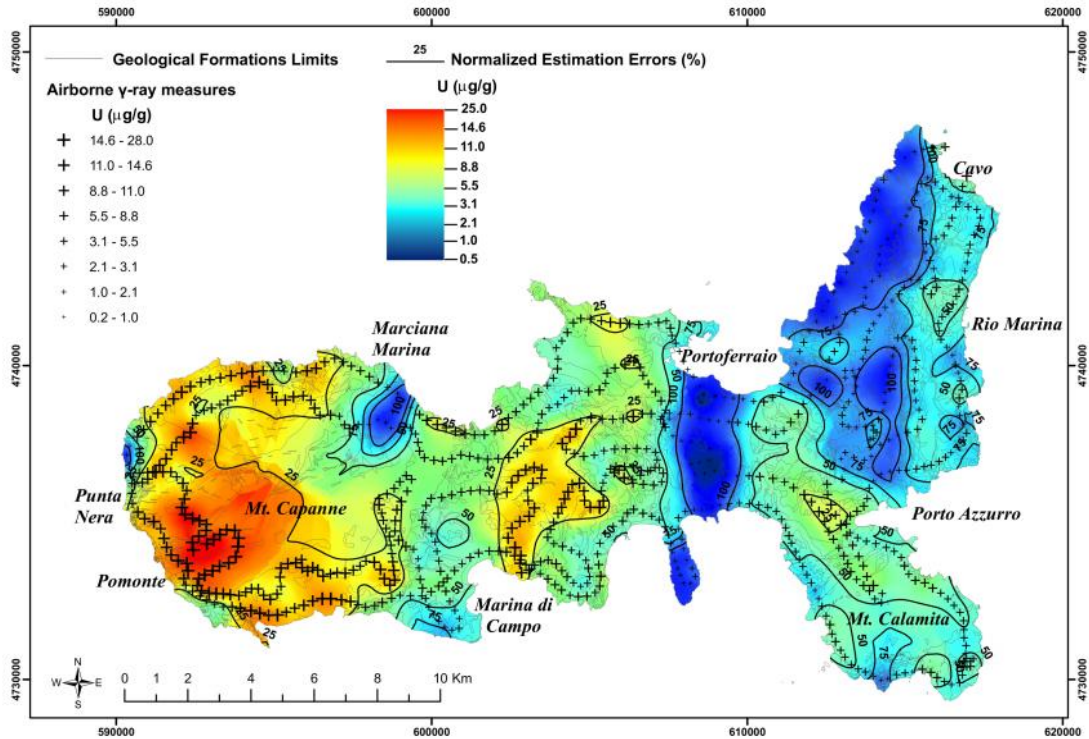

Figure 7. Estimation map of eU (μg/g) abundance and normalized estimation errors.

In the geostatistical approach described above, we faced the problem of correlating a quantitative variable (radioactivity content) to a typical categorical extensive variable (geological map). As a first solution, the standard Gaussian distribution of the secondary variable (Geo1) was chosen in a range of values from $-10^2$ to $10^2$. In order to test possible bias introduced by the choice of the interval of values, we constructed two different distributions in the range of values from 1 to $10^2$ (Geo2) and from 1 to $10^5$ (Geo3). The main results of these tests are summarized in **Table 4** and **Figure 8**; for the sake of simplicity, we only compare here the estimated maps of K abundance. However, the entire procedure for every radioelement combined with the geological parametrical map has been performed. The normalized differences between pairs of maps realized for different casual geological arrays through CCoK interpolations (**Table 4** and **Figure 9**) confirm that the random assignment does not introduce any systematic bias. Moreover, the normalized fluctuations of K abundances estimated by three different models are contained in a range of less than 5%. The quality of the models is not weakened by the assignment of random values to geological categories.

**Table 4**. Descriptive statistics of the CCoK estimation maps of K abundances (unit of measurement: mg/g) using three different parametric classifications of the geological map (Geo1, Geo2, and Geo3), the respective estimation errors maps (NSD), and their algebraic map differences (unit of measurement: %).

| Type | Geological map | Min. | Max. | Mean | Std. Dev. |
|---|---|---|---|---|---|
| CCoK estim. | Geo1 | 0.15 | 48.80 | 19.37 | 0.79 |
| | Geo2 | 0.15 | 48.80 | 19.37 | 0.79 |
| | Geo3 | 0.16 | 48.24 | 19.36 | 0.79 |
| NSD | Geo1 | 0.79 | 187.62 | 27.24 | 19.58 |
| | Geo2 | 0.79 | 217.74 | 27.24 | 19.69 |
| | Geo3 | 6.00 | 255.00 | 27.22 | 19.89 |
| Differ. CCoK | (Geo1-Geo2)/Geo1 | -0.33 | 0.56 | -0.001 | 0.001 |
| | (Geo1-Geo3)/Geo1 | -1.84 | 1.60 | -0.004 | 0.076 |
| | (Geo2-Geo3)/Geo2 | -1.72 | 1.49 | -0.007 | 0.082 |
| Differ. NSD | (Geo1-Geo2)/Geo1 | -44.11 | 91.87 | -1.01 | -0.88 |
| | (Geo1-Geo3)/Geo1 | -85.19 | 90.55 | -0.09 | 1.21 |
| | (Geo2-Geo3)/Geo2 | -34.67 | 49.03 | 0.10 | 0.81 |

The main features of the resulting radiometric maps of abundances for the natural radioelements overlay the prominent geological formations of Elba Island. Indeed, the relevant geological structures defined by the TCs, described in **section 2.1**, can easily be identified by comparing similar abundances of natural radioelements.

The radiometric maps of K, eTh, and eU abundances (**Figures 6**, **7**, and **8**) show high values in the western sector of the island, corresponding to the intrusive granitic complex on Mt. Capanne (indicated as the "CAPa" and "CAPb" geological formations in **Figure** 1). In 19 rock samples of Mt. Capanne pluton reported in **Farina et al. (2010)** the abundances of K, Th, and U are 3.6 ± 0.2 %, 20.8 ± 1.6 μg/g and 8.2 ± 5.1 μg/g respectively. The values match with those estimated in **Figure 5**, **6** and **7**. Although the distributions of radioelements do not distinguish among the three intrusive facies, which are mainly characterized by the widespread occurrence of euhedral K-feldspar megacrystspato, the area with high content of K, Th and U obtained by multivariate analysis follows the contour map of Mt. Capanne pluton reported in Figure 9a in **Farina et al. (2010)**. However, the highest content of K, Th and U are localized in the southwestern part of pluton with the maximum concentration in correspondence of the Pomonte valley, SW-NE oriented that is one of the most prominent morphological lineament of western Elba (**Figure 5**, **6**, and **7**). This is an important tectonic lineament crossing all the Mt. Capanne, abruptly separating two different morphological assets: the north-western part shows rough slopes and deep valleys, whilst the south-eastern one is characterized by gently landscape. The hypothesis of an enrichment of radioelements related to this tectonic lineament should be investigated by further airborne and ground surveys.

As shown in **Figure 5**, **6**, and **7** the geological formations belonging to TC II and TC III have low natural radioelement abundances. The main outcrops are located in the northeastern sector of Elba Island, between Porto Azzurro and Cavo and in the southern part of Portoferraio, where we find peridotites and pillow lavas (indicated as "PRN" and "BRG", **Figure 1**). Finally, low abundance values are found in the area of Punta Nera

Cape at the western edge of the Elba Island, where lithologies belonging to the Ophiolitic Unit crop out (TC IV).

We emphasize that, although we assign a unique number to each geological formation the internal variability of the radiometric data is not biased by the multivariate interpolation. The main evidence of this feature can be observed inside the polygon including Mt. Calamita, which is identified by a unique geological formation "FAFc" (**Figure 1**). We note a clear anomaly of high K abundance in the northeastern sector of the Mt. Calamita promontory, close to Porto Azzurro (**Figure 5**). This anomaly can be geologically explained considering two related factors. The intense tectonization and following fracturation of this sector allowed a significant circulation of magmatic fluids related to the emplacement of monzogranite pluton of La Serra-Porto Azzurro. Moreover, the presence of felsic dykes, metasomatic masses and hydrothermal veins are recently confirmed by **Dini et al. (2008)** and **Mazzarini et al. (2011)**. Although our geological map doesn't report these lithological details, the quality of radiometric survey is such as to identify the location of the felsic dyke swarm. These dykes 30 - 50 cm thick represent the dominant lithology at the mesoscopic scale and their high frequency in FAFc geological formation contributes to increase the gamma-ray signal. These details are not compromised by the multivariate analysis. The spatial extension of high K content validates the geological sketch reported in **Figure 1** by **Dini et al. (2008)**.

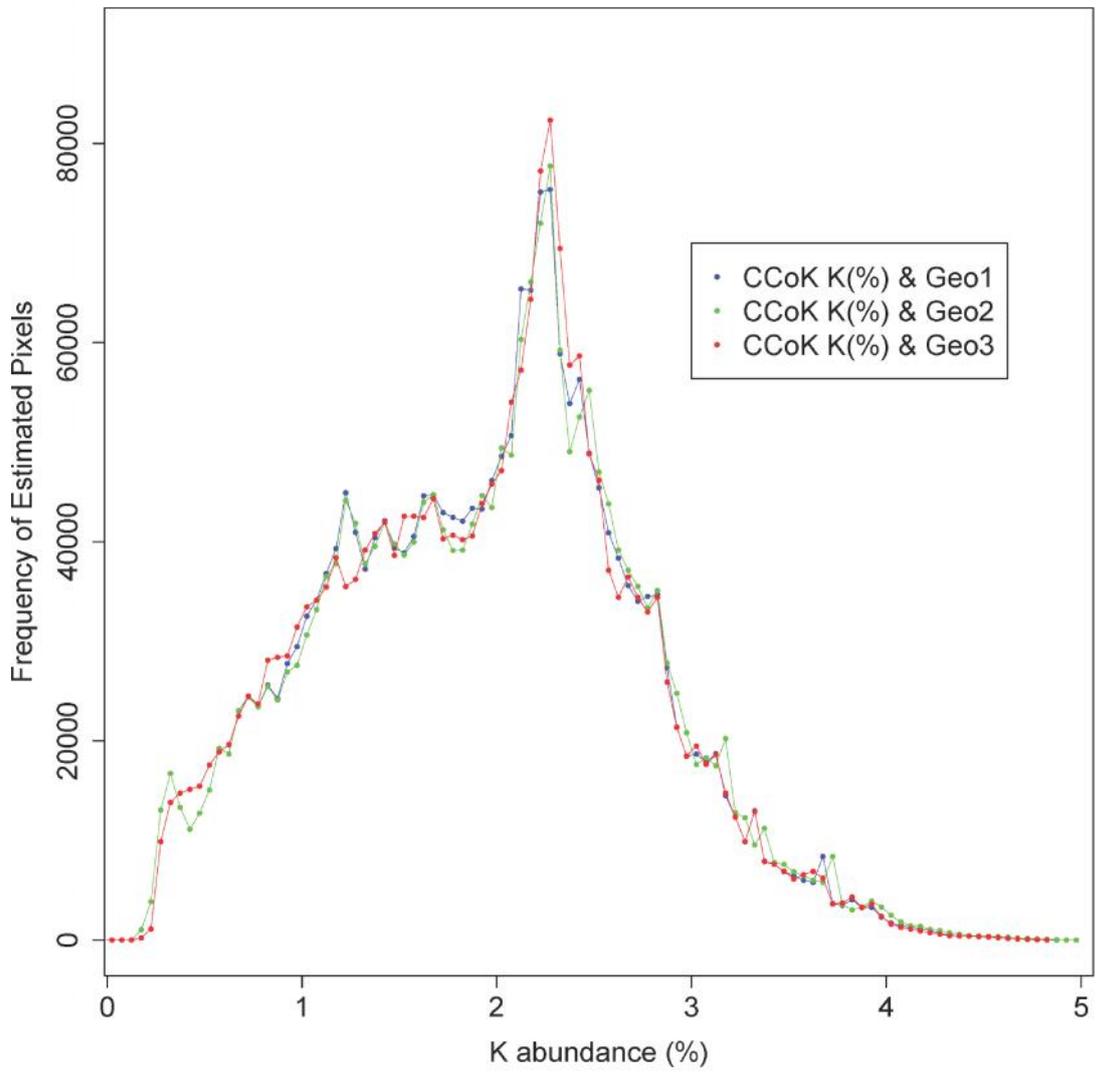

522

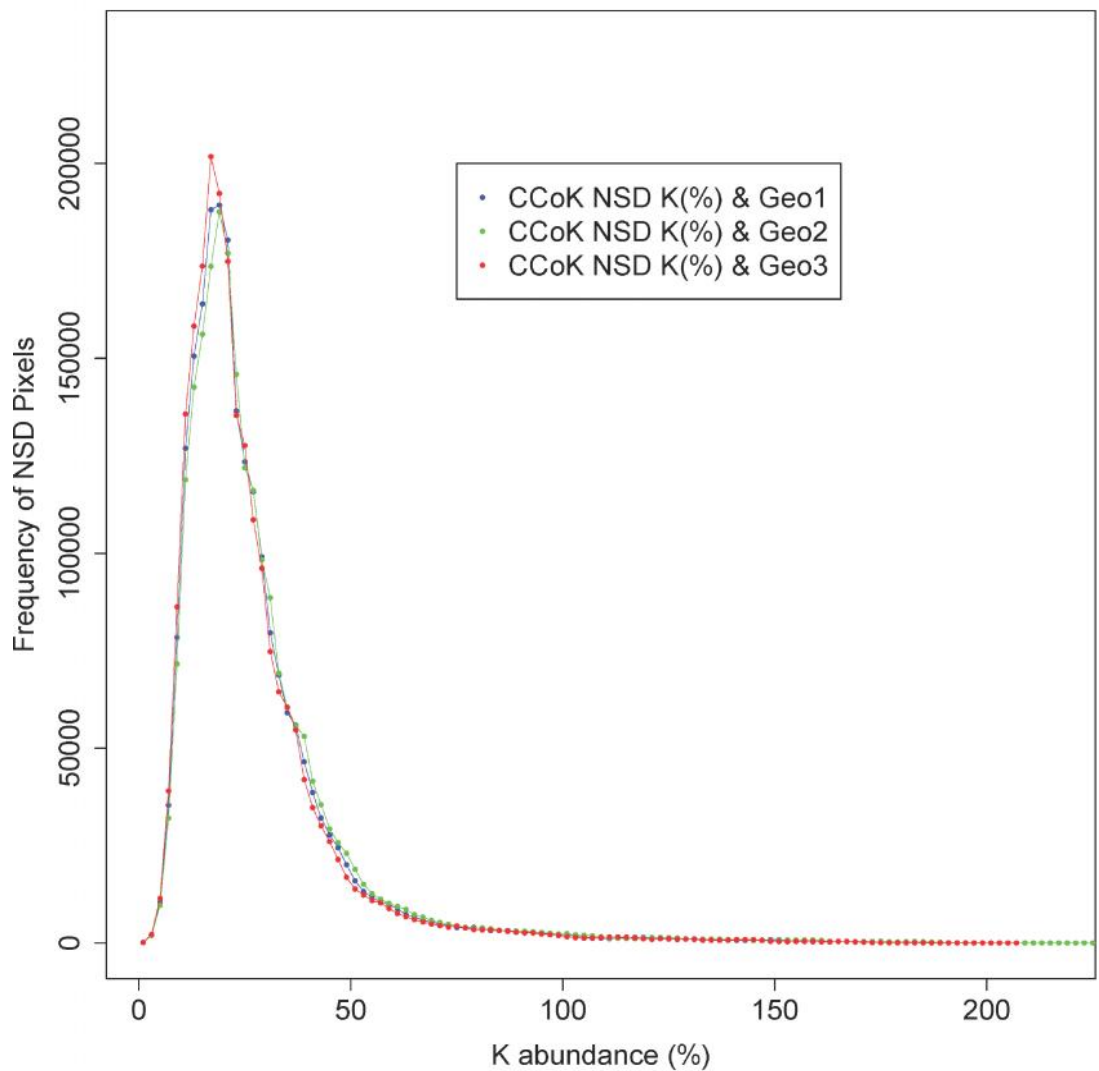

**Figure 8**. a) Frequency distributions of kriged maps of K abundances estimated by CCoK through three different reclassifications of the geological map of Elba Island. b) Frequency distributions of the normalized standard deviation maps (the accuracy of CCoK estimations).

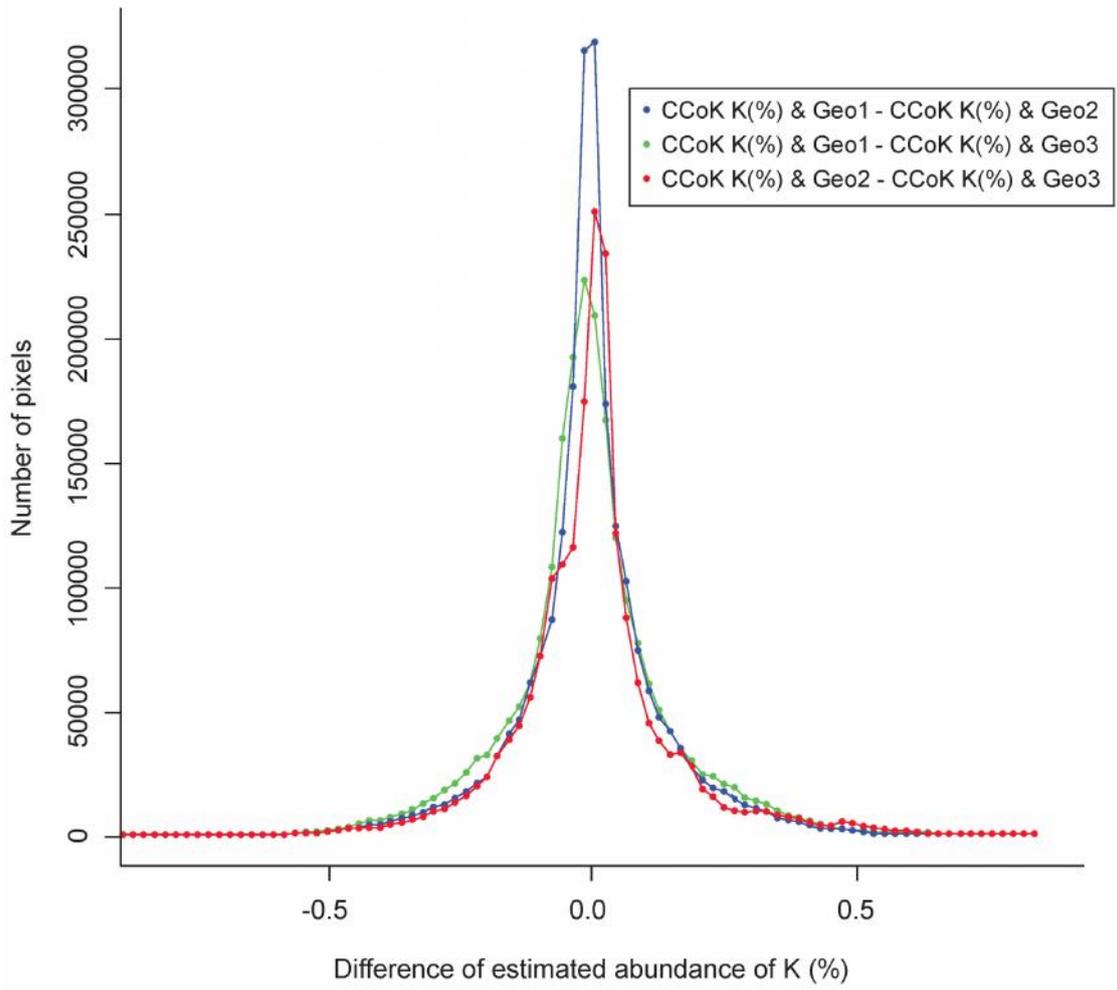

528

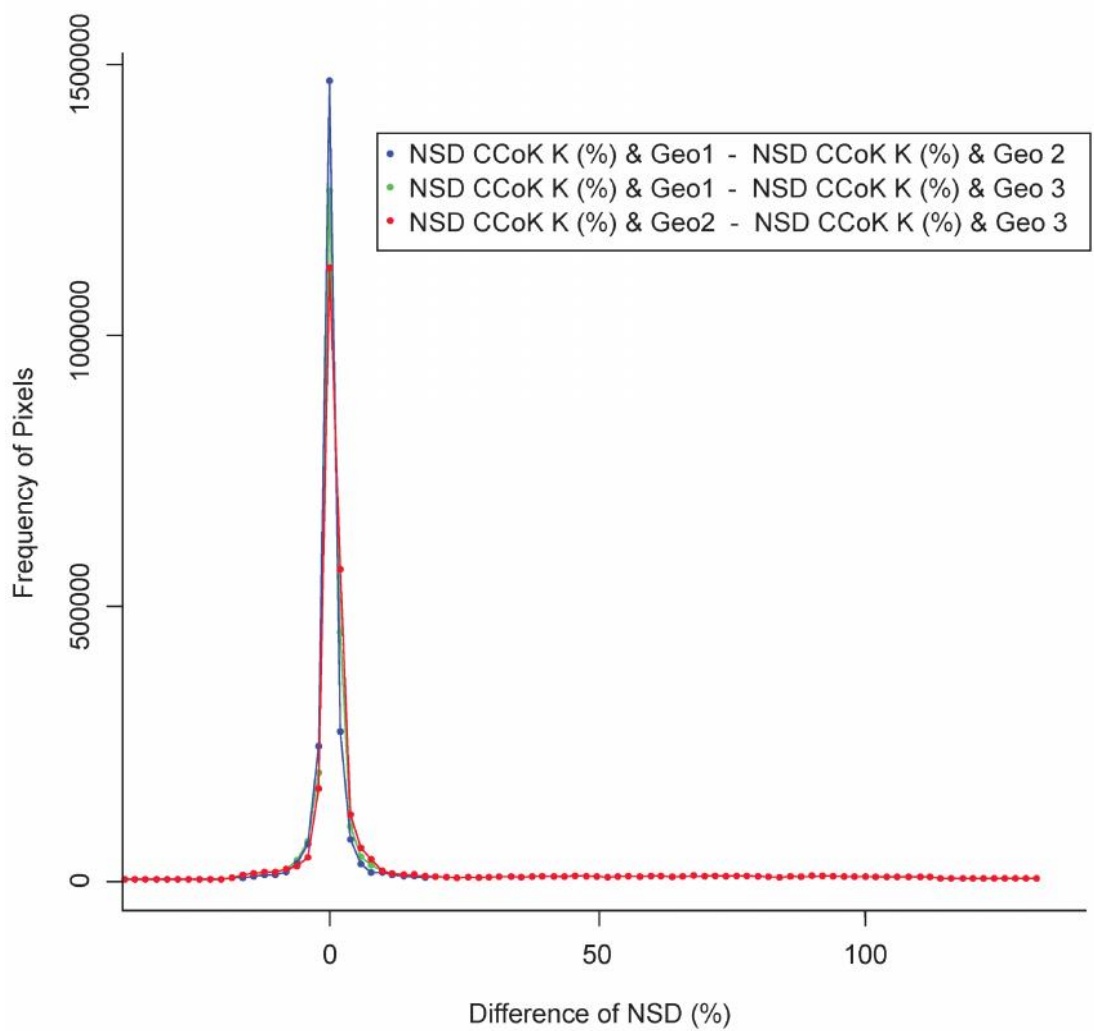

**Figure 9**. a) Frequency distributions of the differences between pairs of kriged maps of K abundances estimated by CCoK through three different reclassifications of geological maps of Elba Island. b) Frequency distributions of the differences between pairs of normalized standard deviation maps.

## 4. Conclusions

In this study we realized the first detailed maps of K, eU, and eTh abundances of Elba Island showing the potential of the multivariate interpolation based on combination of AGRS data and preexisting information contained in the geological map (at scale 1:10,000). We summarize here the main results reached in this study.

- The multivariate analysis technique of collocated cokriging (CCoK) was applied in a non-conventional way, using the well-sampled geology as a quasi-quantitative variable and constraining parameter. This approach gives a well-structured LCMs which show a good spatial co-variation in the omnidirectional coregionalization ESV model. The ESV models show low spatial variability below 600 m, which also corresponds to the radiometric data obtained by partially overlapped spot areas as well as the autocorrelation distance of 1500 m for the three radionuclides. The ESV model of the geology shows a weak variability discontinuity in the first lag, corresponding to the random assignment of quasi-quantitative values of adjacent geological formations, but also a strong spatial relationship up to the first range of autocorrelation. The procedure of the cross-validation of the model yields a mean close to zero for the standardized errors (MSE) and a variance of standardized errors (VSE) close to unity for all groups of variables.
- The CCoK based on the geological constraint was performed by randomly assigning a number to each category of the 73 geological formations. Three different geological quasi-quantitative variable datasets were used, and satisfactory results were achieved by assuring the non-dependency of the model. The normalized fluctuations of three different models are contained in a range of less than 5%.

- Combining the smoothing effects of the probabilistic interpolator (CCoK), and the abrupt discontinuities of the geological map, we observe a distinct correlation between the geological formation and radioactivity content as well as high K, eU and eTh abundances in the intrusive granitic complex on Mt. Capanne and low abundances in the geological formations belonging to TC II, TC III and TC IV.

- Although we assign a unique number to each geological formation, the internal variability of the radiometric data is not biased by the multivariate interpolation. The main evidence of this feature can be observed in the northeastern sector of the geological polygon including Mt. Calamita. A clear anomaly of high K content has confirmed the presence of felsic dykes and hydrothermal veins not reported in our geological map, but recently studied (**Dini et al., 2008**) as a proxy of the high temperature system currently active in the deep portion of Larderello-Travale geothermal field.

## Acknowledgements

The authors are indebted to E. Bellotti, G. Di Carlo, P. Altair, S. De Bianchi, P. Conti and R. Vannucci for useful suggestions and invaluable discussions. This work has been sponsored by the INFN, Cassa di Credito Cooperativo Padova e Rovigo, and funded by Tuscany Region.